\documentstyle[11pt,fleqn]{article}
\catcode`\@=11
\@addtoreset{equation}{section}
\def\theequation{\arabic{section}.\arabic{equation}}
\def\appendix{\renewcommand{\thesection}{\Alph{section}}\setcounter{section}{0}
              \renewcommand{\theequation}
            {\mbox{\Alph{section}.\arabic{equation}}}\setcounter{equation}{0}}
\def\maketitle{\thispagestyle{empty}\setcounter{page}0\newpage
                \renewcommand{\thefootnote}{\arabic{footnote}}
                  \setcounter{footnote}0}
\renewcommand{\thanks}[1]{\renewcommand{\thefootnote}{\fnsymbol{footnote}}
               \footnote{#1}\renewcommand{\thefootnote}{\arabic{footnote}}}
\newcommand{\preprint}[1]{\hfill{\sl preprint - #1}\par\bigskip\par\rm}
\renewcommand{\title}[1]{\begin{center}\Large\bf #1\end{center}\rm\par\bigskip}
\renewcommand{\author}[1]{\begin{center}\Large #1\end{center}}
\newcommand{\address}[1]{\begin{center}\large #1\end{center}}
\def\dip{\smallskip Department of Mathematics, University of Massachusetts,\\
           Amherst, Massachusetts 01003}
\def\Idip{\address{\dip}}
\newcommand{\email}[1]{e-mail: \sl #1@math.umass.edu\rm}
\newcommand{\femail}[1]{\thanks{\email{#1}}}

\def\babs{\hrule\par\begin{description}\item{Abstract: }\it}
\def\eabs{\par\end{description}\hrule\par\medskip\rm}
\renewcommand{\date}[1]{\par\bigskip\par\sl\hfill #1\par\medskip\par\rm}
\newcommand{\ack}[1]{\par\section*{Acknowledgements} #1}
\newcommand{\s}[1]{\section{#1}}
\renewcommand{\ss}[1]{\subsection{#1}}

\def\beq{\begin{eqnarray}}    
\def\eeq{\end{eqnarray}}      
\def\at{\left(}               
\def\ct{\right)}              
\def\R{{\hbox{{\rm I}\kern-.2em\hbox{\rm R}}}}   
\def\H{{\hbox{{\rm I}\kern-.2em\hbox{\rm H}}}}   
\def\N{{\hbox{{\rm I}\kern-.2em\hbox{\rm N}}}}   
\def\C{{\ \hbox{{\rm I}\kern-.6em\hbox{\bf C}}}} 
\def\Z{{\hbox{{\rm Z}\kern-.4em\hbox{\rm Z}}}}   
\renewcommand{\Re}{\mathop{\rm Re}\nolimits}       

\def\ga{\gamma}

\def\Ga{\Gamma}

\def\citen#1{%
\edef\@tempa{\@ignspaftercomma,#1, \@end, }
\edef\@tempa{\expandafter\@ignendcommas\@tempa\@end}%
\if@filesw \immediate \write \@auxout {\string \citation {\@tempa}}\fi
\@tempcntb\m@ne \let\@h@ld\relax \let\@citea\@empty
\@for \@citeb:=\@tempa\do {\@cmpresscites}%
\@h@ld}
\def\@ignspaftercomma#1, {\ifx\@end#1\@empty\else
   #1,\expandafter\@ignspaftercomma\fi}
\def\@ignendcommas,#1,\@end{#1}
\def\@cmpresscites{%
 \expandafter\let \expandafter\@B@citeB \csname b@\@citeb \endcsname
 \ifx\@B@citeB\relax 
    \@h@ld\@citea\@tempcntb\m@ne{\bf ?}%
    \@warning {Citation `\@citeb ' on page \thepage \space undefined}%
 \else
    \@tempcnta\@tempcntb \advance\@tempcnta\@ne
    \setbox\z@\hbox\bgroup 
    \ifnum\z@<0\@B@citeB \relax
       \egroup \@tempcntb\@B@citeB \relax
       \else \egroup \@tempcntb\m@ne \fi
    \ifnum\@tempcnta=\@tempcntb 
       \ifx\@h@ld\relax 
          \edef \@h@ld{\@citea\@B@citeB}%
       \else 
          \edef\@h@ld{\hbox{--}\penalty\@highpenalty \@B@citeB}%
       \fi
    \else   
       \@h@ld \@citea \@B@citeB \let\@h@ld\relax
 \fi\fi%
 \let\@citea\@citepunct
}
\def\@citepunct{,\penalty\@highpenalty\hskip.13em plus.1em minus.1em}%
\def\@citex[#1]#2{\@cite{\citen{#2}}{#1}}%
\def\@cite#1#2{\leavevmode\unskip
  \ifnum\lastpenalty=\z@ \penalty\@highpenalty \fi 
  \ [{\multiply\@highpenalty 3 #1
      \if@tempswa,\penalty\@highpenalty\ #2\fi 
    }]\spacefactor\@m}
\begin{document}
\preprint{}
\title{Zeta-Function Regularization and the Thermodynamic Potential for 
Quantum Fields in Symmetric Spaces}
\author{I. Brevik\thanks{email: Iver.H.Brevik@mtf.ntnu.no},}
\address{Division of Applied Mechanics, Norwegian University of Science
and Technology \\
 N-7034 Trondheim, Norway}
\author{A.A. Bytsenko\thanks{email: abyts@fisica.uel.br\,\,\,\,\,
On leave from Sankt-Petersburg State 
Technical University, Russia},
A.E. Gon\c calves\thanks{email: goncalve@fisica.uel.br}}
\address{Departamento de Fisica, Universidade Estadual de 
Londrina, Caixa Postal 6001, Londrina-Parana, Brazil}\author{and}
\author{F.L. Williams\femail{williams}}\Idip\date{November 1997}

\babs

We calculate a temperature dependent part of the one-loop thermodynamic
potential (and the free energy) for charged massive fields in a general class
of irreducible rank 1 symmetric spaces. Both low- and high-temperature
expansions are derived and the role of non-trivial topology influence on
asymptotic properties of the potential is discussed.

\eabs

\s{Introduction}

The problem of asymptotic expansions of the one-loop 
effective potential in Kaluza-Klein finite temperature theories 
with non-vanishing chemical potential has been studied for a long time by
several authors (see for review Refs. 
\cite{camp90-196-1,eliz94,byts96-266-1}). 
The low- and high-temperature asymptotics in powers of $\beta=1/T$, where 
$T$ is a temperature 
of a system, has been evaluated in terms of integrated heat 
kernel coefficients, related to the scalar and spinor Laplacian 
acting on smooth (compact) $d$-manifold $M^d$ without boundary 
\cite{habe81-46-1497,habe82-23-1852,habe82-25-502,dowk84-1-359,camp90-196-1,
eliz94,byts96-266-1}. An extension of this analysis to a Fermi gas was given
first in Ref. \cite{acto86-256-689}. The boundary conditions for a curved
space in the thermodynamic system with non-vanishing chemical potential 
have been considered in Refs. \cite{dowk89-327-267,camp91-8-529}.
Finite temperature analysis was developed also in manifolds with hyperbolic
spatial sections \cite{byts92-291-26,byts92-7-2669,cogn94-49-5307,eliz94,
byts96-266-1}. The method of the zeta-function regularization in presence of
the multiplicative anomaly for a system of charged bosonic fields with the
non-vanishing chemical potential was reconsidered and actively analyzed
in Ref \cite{eliz97u-71}.  

Only recently topological Casimir energy \cite{will97-38-796}, the one-loop 
effective action, the multiplicative and the conformal anomaly
\cite{byts97u-20,byts97u-19},
associated with product of Laplace type operators acting in rank 1
symmetric spaces, have been analysed.
The goal of this paper is to study the influence of such non-trivial 
topology of manifolds $M^d$ on the one-loop contribution to 
the thermodynamic potential and to the free energy of charged 
massive fields. We consider a general class of irreducible symmetric 
rank 1 Einstein d-manifolds.
Geometric structure on Einstein manifolds $M^d$ is related to an 
Einstein metric $g$, for which the equation $\mbox{Ric}(g)=\Lambda g$ holds, 
where 
$\mbox{Ric}(g)$ is the Ricci tensor and $\Lambda$ is a constant. Trivial 
examples 
of Einstein manifolds are spaces with constant sectional curvature (in 
particular, uniform spaces $\R^d, S^d$ and $\H^d$). 

The contents of the paper are the following. In Sect. 2 we 
review some relevant information on the spectral zeta 
function and the one-loop thermodynamic potential
$\Omega^G(\beta,\mu)$ related to a non-compact simple split rank 1 Lie group
$G$. The low-temperature expansion of the temperature 
dependent part of the potential $\Omega_{\beta}^{G}(\beta,\mu)$ and the free 
energy are calculated in Sect. 3. In
Sect. 4 the explicit form of the high-temperature expansion of 
$\Omega_{\beta}^G(\beta,\mu)$ is presented. We end with some 
conclusions in Sect. 5. Finally the Appendicies A and B contain a summary of
the heat kernel (Appendix A) and the zeta function (Appendix B) properties
relevant to irreducible rank 1 symmetric spaces.

\s{The One-Loop Thermodynamic Potential}

In this section general expressions for the thermodynamic potential and the 
free energy will be derived. For the sake of completeness we shall present the
one-loop contributions of these thermodynamic quantities for the cases in which
the spatial sections are general irreducible rank 1 symmetric spaces (Einstein
manifolds) $X\equiv M^d$ of non-compact type. 
We recall that a Riemannian manifold $(X,g)$ is a 
symmetric space if for any $x\in X$ there exist a group of manifold isometries 
${\Ga}_x$ such that ${\Ga}_x(x)=x$\, and $T_x({\Ga}_x)=-Id_{(T_x X)}$, 
where $T_xX$ is a tangent space. By the way an irreducible rank 1 Riemannian 
symmetric space $(X,g)$ has the form $G/K$, where $G$ is a rank 1 Lie group 
and $K\subset G$ is a maximal compact subgroup (see for example 
\cite{helg62,bess87}).       

We start with the thermodynamic potential for massive charged scalar 
fields with a non-vanishing chemical potential $\mu$ in thermal equilibrium 
at finite temperature in an ultrastatic space-time with spatial sector of the
form $X=G/K$. 
Let $\Gamma\subset G$ be a discrete, co-compact, torsion free subgroup. Let 
$\chi$ be a finite-dimensional unitary representation of $\Gamma$, let 
$\{\lambda_l\}_{l=0}^{\infty}$ be the set of eigenvalues of the
second-order operator of Laplace type ${\cal L}_0=-\Delta_{\Gamma}$ acting on
smooth sections of the vector bundle over $\Gamma\backslash X$ induced
by $\chi$, and let $n_l(\chi)$ denote the multiplicity of $\lambda_l$.

For the ultrastatic space-time with topology $S^1\bigotimes X$, the elliptic 
second order differential operator $L(\mu)$ is a matrix valued 
operator acting on the real and imaginary part of the complex scalar 
charged field. It has the form $L(\mu)
=\mbox{diag} \at -(\partial_{\tau}-e\mu)^2+{\cal L}, -(\partial_{\tau}+e\mu)^2
+{\cal L} \ct \equiv \mbox{diag}({\cal O}_+,{\cal O}_-)$; to simplify the 
calculation we take $e=1$, where $e$ is an elementary charge. 
The operators ${\cal O}_{\pm}$ are not hermitian; in fact they are  normal 
and their eigenvalues are complex and read
$$
\left(\frac{2\pi n}{\beta}\pm i\mu\right)^2+\lambda_l+b;\,\,\,
n\in \Z
\mbox{.}
\eqno{(2.1)}
$$
In Eq. (2.1) $b$ is an arbitrary constant (an endomorphism of the vector 
bundle over $\Ga\backslash X$).
We shall need further a suitable regularization of the determinant of an
elliptic differential operator, and we shall make choice of the zeta-function 
regularization. The zeta function associated with the operator 
${\cal L}\equiv {\cal L}_0+b$ has the form

$$\zeta_{\Ga}(s|{\cal L})=\sum_ln_l(\chi)\{\lambda_l+b\}^{-s}\mbox{;}
\eqno{(2.2)}
$$
where $\zeta_{\Ga}(s|{\cal L})$ is a well-defined analytic function for
$\Re s >\mbox{dim}(X)/2$, and it can be analytically continued to a 
meromorphic function on the complex plane $\C$, regular at $s=0$. 

The canonical partition function can be written as follows 
$$
\log 
Z(\beta,\mu)=-\beta\Omega^{G}(\beta,\mu)=-S_c[\phi_c,g]-\frac{1}{2}
\log\mbox{det}[L(\mu)]
\mbox{,}
\eqno{(2.3)}
$$ 
where $\phi_c$ is a solution, which extremizes the classical action $S_c
[\phi_c,g]$; the Eq. (2.3) defines the thermodynamic potential 
$\Omega^{G}(\beta,\mu)$. 
If one makes use of zeta-function regularization, one obtains 
${\zeta}_{\Ga}'\left(0|{\cal O}_-\right)={\zeta}_{\Ga}'\left(0|{\cal O}_+
\right)$ and   
$$
\Omega^{G}(\beta,\mu)=\frac{1}{\beta}S_c\left[\phi_c,g\right]-
\frac{1}{\beta}{\zeta}_{\Ga}'\left(0|{\cal O}_+\right)-\frac{1}{2 \beta}
{\cal A}({\cal O}_+,{\cal O}_-)
\mbox{,}
\eqno{(2.4)}
$$
where
${\cal A}({\cal O}_+,{\cal O}_-)$ is the related multiplicative anomaly
\cite{eliz97u-71}. The explicit form of the ${\cal A}({\cal O}_+,{\cal O}_-)$
associated with spaces $X$ listed in Eq. (A.3) of the Appendix A, can be
found in Ref. \cite{byts97u-20}.

Using the Mellin representation for the zeta function one can obtain useful
formulae for the non trivial temperature dependent part $\Omega_{\beta}^{G}
(\beta,\mu)$
of the thermodynamic potential (see for detail Ref. \cite{byts96-266-1,
eliz97u-71})
$$
\Omega_{\beta}^{G}(\beta,\mu)\equiv \Omega^{G}(\beta,\mu)-\Omega_0^{G}
-\frac{1}{2\beta}{\cal A}({\cal O}_+,{\cal O}_-)=
-\frac{1}{\pi}\sum_{\nu=1}^\infty\int_{\R}e^{i\nu\beta t}
{\zeta}_{\Ga}'\left(0\left| \cal{L}\right. +
[t +i\mu]^2 \right)dt
$$

$$
=-\frac{1}{\sqrt{\pi}}\sum_{\nu=1}^\infty\cosh(\nu\beta\mu)\int_0^\infty 
t^{-3/2}e^{-\nu^2\beta^2/4t}\omega_{\Ga}(t;b,\chi)dt
\eqno{(2.5)}
$$

$$
=-\frac{1}{\pi i}\sum_{\nu=0}^\infty\frac{\mu^{2\nu}}{(2\nu)!}
\int_{\Re s=c}\zeta_R(s)\Gamma(s+2\nu-1)\zeta_{\Ga}
\left(\frac{s+2\nu-1}{2}\left|{\cal L}\right.\right)\beta^{-s}ds
\mbox{,}
\eqno{(2.6)}
$$
where
$$
\Omega_0^{G} = \frac{1}{\beta}S_c\left[\phi_c,g\right]+\xi^{(r)}
\left(-\frac{1}{2}|{\cal L}\right)
\mbox{,}
\eqno{(2.7)}
$$

$$
\xi^{(r)}
\left(-\frac{1}{2}|{\cal L}\right)=PP\zeta_{\Ga}(-\frac{1}{2}|{\cal L})+
(2-2\log 2)\mbox{Res} \zeta_{\Ga}(-\frac{1}{2}|{\cal L})
\mbox{.}
\eqno{(2.8)}
$$
In Eq. (2.5) $\omega_{\Ga}(t;b,\chi)$ is the heat kernel of an operator 
${\cal L}$ (see Eqs. (A.1), (A.2), (A.6) and (A.8) of the Appendix A); the 
symbols $(PP)$ and 
$(\mbox{Res})$ in Eq. (2.8) stand for the finite part and the residue of the 
function at the special point respectively. The formulae (2.5), (2.6) are 
valid for a charged scalar field. In the case of a neutral scalar field we have
to multiply all results by the factor $1/2$.

Different representations of the temperature dependent part of the 
thermodynamic potential can be obtained by means of Eqs. (2.5) and (2.6). 
In fact, using Eq. (A.6), (A.7) for the heat kernel in the formula (2.5) we 
obtain
$$
\Omega_\beta^G(\beta,\mu)=-\frac{1}{\sqrt{\pi}}\sum_{\nu=1}^{\infty}\cosh 
(\nu\beta\mu)
\int_0^{\infty}t^{-\frac{3}{2}}e^{-\frac{\nu^2\beta^2}{4t}}\left[V\int_{\R}
e^{-(r^2+\alpha^2)t}|C(r)|^{-2}dr+\theta_{\Ga}(t;b,\chi)\right]dt
\mbox{.}
\eqno{(2.9)}
$$
For the sake of generality we shall set $b+\rho_0^2\equiv\alpha^2$, where
$\alpha$ is an arbitrary constant. Note that 
$\alpha^2=\rho_0^2$ and $\alpha^2=0$ correspond to the massless and conformal
coupling case respectively.
Taking into account an integral representation for the Mac Donald functions 
$K_{\nu}(z)$: 
$$
K_{\nu}(z)=\frac{1}{2}\left(\frac{z}{2}\right)^{\nu}\int_{0}^{\infty}
e^{-t-\frac{z^2}{4t}}t^{-\nu-1}dt,\,\,\,\,\,\,\,\,\,\,
\left(|\mbox{arg} z|<\frac{\pi}{2}\,\,\,\,\,
\mbox{and}\,\,\, \Re z^2>0\right),\,\,\,\,\,\,\,\,\,
\mbox{(2.10)}
$$
we get
$$
\Omega_{\beta}^G(\beta,\mu)=-2\sqrt{\frac{2}{\pi}}\sum_{\nu=1}^{\infty}
\cosh (\nu\beta\mu)\left[V\int_{\R}|C(r)|^{-2}\left(\frac{\sqrt{r^2+\alpha^2}}
{\nu\beta}\right)^{\frac{1}{2}}K_{\frac{1}{2}}\left(\nu\beta\sqrt{r^2+\alpha^2}
\right)dr\right.
$$
$$
\left.+\frac{1}{\sqrt{2\pi}}\sum_{\ga\in C_{\Ga}-\{1\}}\chi(\ga)t_{\ga}j(\ga)
^{-1}C(\ga)\frac{\alpha}{\sqrt{\nu^2\beta^2+t_{\ga}^2}}K_1\left(\alpha
\sqrt{\nu^2\beta^2+t_{\ga}^2}\right)\right]
\mbox{.}
\eqno{(2.11)}
$$

\s{The Low-Temperature Expansion}

\ss{The Thermodynamic Potential}

The formulae (2.5) and (2.6) are useful for the low- and
high-temperature expansion of the thermodynamic quantity \cite{byts96-266-1}. 
Indeed, in order to specialize the Eq. (2.5) for the low-temperature case let 
us recall the asymptotic of the Mac Donald functions for real values $z$ and 
$\nu$, namely
$$
K_{\nu}(z\mapsto\infty)\simeq\sqrt{\frac{\pi}{2z}}e^{-z}\sum_{k=0}^
{\infty}\frac{\Ga(\nu+k+\frac{1}{2})}{\Ga(k+1)\Ga(\nu-k+\frac{1}{2})}
(2z)^{-k}
\mbox{.}
\eqno{(3.1)}
$$
As a result we have
$$
\Omega_{\beta}^G(\beta\mapsto\infty,\mu)\simeq -\sum_{\nu=1}^{\infty}
\left[\frac{V}{\nu\beta}\int_{\R}|C(r)|^{-2}
e^{-\nu\beta(\sqrt{r^2+\alpha^2}-|\mu|)}dr\right.
$$

$$
\left.+\sqrt{\frac{\alpha}{2\pi}}\sum_{\ga\in C_{\Ga}-\{1\}}\sum_{k=0}^{\infty}
\frac{\chi(\ga)t_{\ga}j(\ga)^{-1}C(\ga)}{(2\alpha)^k(\nu^2\beta^2+t_{\ga}^2)^
{\frac{k}{2}+\frac{3}{4}}}\frac{\Ga(k+\frac{3}{2})}{\Ga(k+1)\Ga(\frac{3}
{2}-k)} e^{-\alpha\sqrt{\nu^2\beta^2+t_{\ga}^2}+\nu\beta|\mu|}\right]
\mbox{.}
\eqno{(3.2)}
$$
Finally using the explicit form of the Harish-Chandra-Plancherel measure
(B.5) in Eq. (3.2) after straightforward calculation we get

$$
\Omega_{\beta}^G(\beta\mapsto\infty,\mu)\simeq 
\frac{2\pi C_GV}{\beta}\sum_{\nu=1}^
{\infty}\sum_{l=0}^{\frac{d}{2}-1}\frac{a_{2l}}{\nu}
\int_{\R}r^{2l+1}e^{-\nu\beta\left(\sqrt{r^2+\alpha^2}-|\mu|\right)}
{\cal R}_G(r)dr
$$

$$
-\sqrt{\frac{\alpha}{2\pi}}\sum_{\nu=1}^{\infty}\sum_{\ga\in C_{\Ga}
-\{1\}}\sum_{k=0}^{\infty}\frac{\chi(\ga)t_{\ga}
j(\ga)^{-1}C(\ga)}{(2\alpha)^k(\nu^2\beta^2+t_{\ga}^2)^{\frac{k}{2}+\frac{3}
{4}}}\frac{\Ga(k+\frac{3}{2})}{\Ga(k+1)\Ga(\frac{3}
{2}-k)} e^{-\alpha\sqrt{\nu^2\beta^2+t_{\ga}^2}+\nu\beta|\mu|}
\mbox{,}
\eqno{(3.3)}
$$
where
$$
{\cal R}_G(r)=\left[ \begin{array}{ll}\left(1+e^{2\pi r}\right)^{-1}, \hspace
{0.8cm} G=SO_1(2n,1),\\
\left(1+e^{\pi r}\right)^{-1}, \hspace{1.0cm}  G=SP(m,1),\,\, F_{4(-20)},\,\,
SU(m,1)\,\,\,\mbox{for odd}\,\, m,\\
\left(1-e^{\pi r}\right)^{-1}, \hspace{1.0cm} G=SU(m,1)\,\,
\mbox{for even}\,\, m,\\
-(2r)^{-1}, \hspace{1.5cm} G=SO_1(2n+1,1).\\
\end{array} \right]
\eqno{(3.4)}
$$
The leading terms come from ``topological'' part of the thermodynamic
potential, related to the function $\theta_{\Ga}(t;b,\chi)$, which is determed
by Eq. (A.8).

\ss{The Free Energy}

The one-loop free energy can be derived from the thermodynamic potential (3.3)
in the limit $\mu\mapsto 0$. Thus the formulae for the free energy can be 
considered as a particular case of expressions given for thermodynamic
potentials. The low-temperature contribution to the one-loop free 
energy has the form 
$$
\Omega_{\beta}^G(\beta\mapsto\infty,0)\simeq \frac{2A}{\beta}
\sum_{l=0}^{\frac{d}{2}-1}a_{2l}
\int_{\R}r^{2l+1}  
\left[\beta\sqrt{r^2+\alpha^2}-\mbox{log}\left(e^{\beta\sqrt{r^2+\alpha^2}}
-1\right)\right]{\cal R}_G(r)dr
$$
$$
-\frac{1}{\sqrt{2\pi}}\sum_{\nu=1}^{\infty}\sum_{k=0}^{\infty}
\sum_{\ga\in C_{\Ga}-\{1\}}
\frac{\chi(\ga)t_{\ga}
j(\ga)^{-1}C(\ga)}{(2\alpha)^{k-\frac{1}{2}}(\nu^2\beta^2+t_{\ga}^2)^
{\frac{k}{2}+\frac{3}{4}}}\frac{\Ga(k+\frac{3}{2})}{\Ga(k+1)\Ga(\frac{3}
{2}-k)} e^{-\alpha\sqrt{\nu^2\beta^2+t_{\ga}^2}}
\mbox{,}
\eqno{(3.5)}
$$
where $A=\pi C_GV$.

\s{The High-Temperature Expansion}

For the high-temperature expansion it is convenient to use the Mellin-Barnes 
representation (2.6) and integrate it on a closed path enclosing a suitable 
number of poles. To carry out the integration first of all we shall consider 
the simplest case of $G=SO_1(2n+1,1)$ in (B.5).  

\ss{The Group ${\bf G=SO_1(2n+1,1)}$}

Taking into account Eq. (2.6) we recall that the zeta
function in a $(2n+1)$-dimensional smooth manifold without boundary has 
simple poles at the points $s=(2n+1)/2-k,\,\,k\in\N$,\,\,
\cite{seel67-10-172}. The Riemann zeta function $\zeta_R(s)$ has a simple pole
at $s=1$ and simple zeros at all the negative even numbers while the function
$\Ga(s)$ has simple poles at $s=-k,\,\,k\in\N$. The temperature dependent part 
of the thermodynamic potential can be written as follows
$$
\Omega_{\beta}^{SO_1(2n+1,1)}(\beta\mapsto 0,\mu)=-\frac{1}{\pi i}\sum_{\nu=0}^
{\infty}\frac{\mu^{2\nu}}
{(2\nu)!}\int_{\Re s=c}{\cal F}^{SO_1(2n+1,1)}(s;\nu,\beta)ds
\mbox{,}
\eqno{(4.1)}
$$
where
$$
{\cal F}^{SO_1(2n+1,1)}(s;\nu,\beta)=2^{s+2\nu-2}\pi^{-\frac{1}{2}}\Ga\left(
\frac{s}{2}+\nu\right)\zeta_R(s)\beta^{-s}
$$
$$
\times\left[
A\sum_{j=0}^na_{2j}\alpha^{2j-2\nu -s+2}\Ga\left(j+\frac{1}{2}\right)
\Ga\left(\frac{s}{2}+\nu-j-1\right)
+T_{\Ga}\left(\frac{s+2\nu-1}{2};\alpha,\chi\right)\right]
\mbox{.}
\eqno{(4.2)}
$$
The meromorphic integrand ${\cal F}^{SO_1(2n+1,1)}(s;\nu,\beta)$ 
has simple poles at 
the points $s=1\,\,\,(\nu\in\N),\,\,s=-2(\nu+k)\,\,\,(k\in\N,\,\,
k=-1,...,-n-1)$ and $s=0\,\,\,(\nu=1,2,...,n+1)$. Moreover for $\nu=0$ at 
$s=0$ we have simple and double poles.
Thus choosing a contour of integration in the left half-plane we obtain the
high-tempertature expansion in the form
$$
\Omega_{\beta}^{SO_1(2n+1,1)}(\beta\mapsto 0,\mu)=-\frac{A}{\sqrt{\pi}}
\left\{\sum_{k=1}^{n+1}
\sum_{\nu=0}^{k-1}\sum_{j\geq k-1}^n(-1)^{j+1-k}2^{2k-1}a_{2j}\right.
$$
$$
\left. \times\frac{\mu^{2\nu}\Ga(k)\Ga(j+\frac{1}{2})}{(2\nu)!\Ga(j+2-k)}
\zeta_R(2k-2\nu)
\alpha^{2j-2k+2}\beta^{2(\nu-k)}\right.
$$

$$
+\left[\sum_{\nu=0}^{\infty}\sum_{j=0}^n
2^{2\nu-1}a_{2j}\frac{\mu^{2\nu}\Ga(\nu+\frac{1}{2})}{(2\nu)!}
\Ga(j+\frac{1}{2})\Ga(\nu-j-\frac{1}{2})\alpha^{2j-2\nu +1}\right.
$$

$$
\left.+\int_{0}^{\infty}\Psi_{\Ga}\left(t+\rho_0+\alpha;\chi\right)dt
\right]\beta^{-1}
$$

$$
\left.+\sum_{j=0}^n(-1)^{j+1}
a_{2j}\frac{\Ga(j+\frac{1}{2})}{\Ga(j+2)}\alpha^{2j+2}\left[\ga_{E}+
\log\left(\frac{\alpha^2\beta}{2}\right)\right]\right.
$$

$$
\left.+\sum_{\nu=1}^{n+1}\sum_{j\geq \nu-1}^n
(-1)^{j-\nu}2^{2\nu-2}a_{2j}\frac{\mu^{2\nu}\Ga(\nu)\Ga(j+\frac{1}{2})}
{(2\nu)!\Ga(j+2-\nu)}\alpha^{2j+2}\right\}
$$

$$
+\frac{1}{2\pi}\int_{0}^{\infty}\Psi_{\Ga}\left(t+\rho_0+\alpha;\chi\right)
\left(2\alpha t +t^2\right)^{\frac{1}{2}}dt + {\cal O}(\beta^2)
\mbox{,}
\eqno(4.3)
$$
where $\ga_{E}$ is the Euler constant,
$$
{\cal O}(\beta^2)=-\frac{A}{\sqrt{\pi}}\sum_{\scriptstyle \nu,k=0,
\atop\scriptstyle n+k\neq 0}^{\infty}\sum_{j=0}^n
(-1)^{k+j+1}a_{2j}\frac{\mu^{2\nu}\Ga(j+\frac{1}{2})}
{2^{2k+1}(2\nu)!\Ga(j+k+2)}
$$
$$
\times \Ga(-k)\zeta_R(-2k-2\nu)
\alpha^{2j+2k+2}
\beta^{2(\nu+k)}
\mbox{.}
\eqno{(4.4)}
$$
At high temperature and even for zero chemical
potential ``topological'' terms give a contribution to the potential. The 
high-temperature expansion (4.3) in principal looks quite similar to the one 
obtained in \cite{byts96-266-1} for $X=\H^3/\Ga$.

For the case of minimally coupled scalar field in manifolds $S^1\otimes 
M^d$
we have $\alpha^2=m^2+\rho_0^2$, where $b=m^2$,\,  $m$ is a mass of field. 
For example, when $n=1,\,\,(d=3)$, the leading term of the Laurent series (4.3)
has the form $-4Aa_{21}\pi^4/(90\beta^4)$, which is a known result 
\cite{dowk84-1-359,dowk89-327-267}.

\ss{The Group ${\bf G \neq SO_1(2n+1,1),\,\,SU(d/2,1)}$}

For $G=SO_1(2n,1), SU(2p+1), SP(m,1), F_{4(-20)}$ the integrand in
Eq. (2.6) has the form
$$
{\cal F}^{G}(s;\nu,\beta)=\zeta_R(s)\beta^{-s}\left[\frac{Aa(G)\Ga(s+2\nu-1)}
{2}W\left(\frac{s+2\nu-1}{2};\alpha^2,a(G)\right)\right.
$$
$$
\left. +\frac{2^{s+2\nu-2}}{\sqrt{\pi}}
\Ga\left(\frac{s}{2}+\nu\right)T_{\Ga}\left(\frac{s+2\nu-1}{2};\alpha,
\chi\right)
\right]
\mbox{,}
\eqno{(4.5)}
$$
where $W(s;\alpha,a(G))$ is given by Eq. (B.7). Therefore the 
temperature dependent part of the thermodynamic potential is
$$
\Omega_{\beta}^{G}(\beta\mapsto 0,\mu)=-a(G)A\
\sum_{\nu=0}^{\infty}\sum_{m,j=0}^{\frac{d}{2}-1}a_j
\frac{\mu^{2\nu}}{(2\nu)!}
j!\Ga(2m+2)\zeta_R(2m+3-2\nu)
$$

$$
\times\sum_{l=m}^j\frac{{\cal K}_{j-l}
(m-l;\alpha^2,a(G))}{(j-l)!}\prod_{\scriptstyle q=0, \atop\scriptstyle  
q\neq m}^l(m-q)^{-1}\beta^{2\nu-2m-3}
$$

$$
-a(G)A\left[-W(0;\alpha^2,a(G))\mbox{log}\beta+W'(0;
\alpha^2,a(G))\right]\beta^{-1}
$$

$$
+a(G)A\sum_{\nu=1}^{\frac{d}{2}}\frac{\mu^{2\nu}\Ga(2\nu)}
{(2\nu)!}
\left[U(\nu;\alpha^2,a(G))+\ga_{E} V(\nu;\alpha^2,a(G))\right.
$$
$$
\left. -V(\nu;\alpha^2,a(G))
\mbox{log}\beta+\psi'(2\nu)
V(\nu;\alpha^2,a(G))\right]\beta^{-1}
+\beta^{-1}\int_{0}^{\infty}\Psi_{\Ga}(t+\rho_0+\alpha;\chi)dt
$$

$$
+\frac{1}{2\pi}\int_{0}^{\infty}\Psi_{\Ga}(t+\rho_0+\alpha;\chi)(2\alpha t+t^2)
^{\frac{1}{2}}dt+{\cal O}(\beta)
\mbox{,}
\eqno{(4.6)}
$$
where $\psi(s)\equiv \Ga'(s)/\Ga(s)$, 
$$
{\cal O}(\beta)=-a(G)A\sum_{\nu=0}^{\infty}\sum_{k=1}^{\infty}
\frac{\mu^{2\nu}}{(2\nu)!
(2k)!}\zeta_R(1-2\nu-2k)W(-k;\alpha^2,a(G))\beta^{2\nu+2k-1}
\mbox{,}
\eqno{(4.7)}
$$

$$
U(s;\alpha^2,a(G))=\sum_{j=0}^{\frac{d}{2}-1}\sum_{\scriptstyle l=0,
\atop\scriptstyle l<n-1}^ja_{2j}j!
\frac{{\cal K}_{j-l}(s-l-1;\alpha^2,a(G))}
{(j-l)!(s-1)(s-2)...(s-(l+1))}
\mbox{,}
$$
$$
V(s;\alpha^2,a(G))=\sum_{j=0}^{\frac{d}{2}-1}\sum_{l\geq n-1}^j
a_{2j}j!\frac{{\cal K}_{j-l}(s-l-1;\alpha^2,a(G))}
{(j-l)!(s-1)(s-2)...(s-(l+1))}
\mbox{,}
\eqno{(4.8)}
$$
and
$$
W'(0;\alpha^2,a(G))=\sum_{j=0}^{\frac{d}{2}-1}\sum_{l=0}^j\frac
{a_{2j}j!(-1)^{l+1}}{(j-l)!(l+1)!}\left[{\cal K}_{j-l}'(-l-1;\alpha^2,
a(G))
\right.
$$

$$
\left.+\frac{1}{2}{\cal K}_{j-l}(-l-1;\alpha^2,a(G))
\sum_{m=1}^{l+1}\frac{1}{m}
\right]
\mbox{.}
\eqno{(4.9)}
$$

\ss{The Group ${\bf G=SU(p,1)}$}

Finally for $G=SU(p,1),\,\,d=2p$, one gets
$$
\Omega_{\beta}^{SU(p,1)}(\beta\mapsto 0,\mu)=\Omega_{\beta}^{G}
(\beta\mapsto 0,\mu)
-2A\sum_{j=0}^{p-1}a_{2j}\left[\sum_{\nu=1}^{\infty}\frac{\mu^{2\nu}
\Ga(2\nu)}{(2\nu)!}{\cal J}_j(\nu;\alpha^2,\frac{\pi}{2})+{\cal J}_j'
(0;\alpha^2,
\frac{\pi}{2})\right]\beta^{-1}
$$

$$
-2A\sum_{\nu=1}^{\infty}\sum_{k=0}^{\infty}\sum_{j=0}^{p-1}
a_{2j}\frac{\mu^{2\nu}\zeta_R (1-2\nu-2k)}{(2\nu)!(2k)!}{\cal J}_j
(-k;\alpha^2,\frac{\pi}{2})\beta^{2\nu+2k-1}
\mbox{.}
\eqno{(4.10)}
$$
In Eq. (4.10) $G\neq SO_1(2n+1,1),\,SU(d/2,1)$ and $a(G)=\pi/2$ has 
been choosen.

\s{Conclusions}

In this paper an extension of previous results to the case in which 
chemical potential for quantum fields in irreducible symmetric spaces of rank 
1 is present has been proposed. In the case of low and high temperature we
obtain a generalisation of the results discussed in \cite{camp90-196-1,eliz94,
byts96-266-1}. 

For the vector (spin 1) field the Hodge-de Rham operator 
$(d\delta+\delta d)$ acting on the exact 
one-forms is associated with the massless operator $[-\nabla^{\mu}
\nabla^{\nu}+(d-1)]g_{\mu\nu}$. The eigenvalues of the operator are 
$\lambda_l^2+(\rho_0-1)^2$ and for the Proca field of mass $m$ we find 
$\alpha^2=m^2+(\rho_0-1)^2$. 

Our results can also be extended to spin $1/2$ (fermion) field, for which spin
structure on a manifold have to be taken into account. Note that different
spin structures are parametrized by the first cohomology group $H^1(X;\Z_2)$.
Asymptotic expansions for spin $1/2$ field can be obtained using the
relation $\Omega_F(\beta,\mu)=2\Omega_B(2\beta,\mu)-\Omega_B(\beta,\mu)$
\cite{dowk89-327-267,byts96-266-1}, where symbol $F\,\, (B)$ stands for fermion
(boson) degree of freedom.

We hope that proposed analysis of 
the one-loop thermodynamic properties of the potential will be interesting in 
view of future applications to concrete problems in
quantum field theory at finite temperature, in quantum gravity (see 
Ref. \cite{buch92}), in multidimensional cosmological models, and in 
mathematical applications as well.

\ack{We thank Prof. S. Zerbini for useful discussions. A.A. Bytsenko wishes 
to thank CNPq and the Department of Physics of 
Londrina University for financial support and kind hospitality. The research 
of A.A. Bytsenko was supported in part by Russian Foundation for Fundamental 
Research grant No. 95-02-03568-a and 98-02-18380 and by Russian Universities 
grant No.~95-0-6.4-1.}

\appendix

\s{The Heat Kernel}

One can define the
heat kernel of the elliptic operator ${\cal L}$ by
$$\omega_{\Gamma}(t;b,\chi)\equiv\mbox{Tr}\left(e^{-t{\cal L}}\right)
=\frac{-1}{2\pi i}\mbox{Tr}\int_{{\cal C}_0}dze^{-zt}(z-{\cal L})^{-1}
\mbox{,}
\eqno{(A.1)}
$$
where ${\cal C}_0$ is an arc in the complex plane $\C$. By standard results 
in operator theory there exist $\epsilon,\delta >0$ such that for $0<t<\delta$ 
the heat kernel expansion holds
$$
\omega_{\Gamma}(t;b,\chi)=\sum_{l=0}^{\infty}n_l(\chi)e^{-(\lambda_l+b)t}=
\sum_{0\leq l\leq l_0} a_l({\cal L})t^{-l}+ O(t^\epsilon)
\mbox{.}
\eqno{(A.2)}
$$

The following representations of $X$ up to local isomorphism can be choosen
$$
X=\left[ \begin{array}{ll}SO_1(n,1)/SO(n)\,\,\,\,\,\,\,\,\,\,\,\,\,\,\,\,\,
\,\,\,\,\,\,\,\,\,\,\,\,\,\,\,\,\,\,\,\,\,(I) \\
SU(n,1)/U(n)\,\,\,\,\,\,\,\,\,\,\,\,\,\,\,\,\,\,\,\,\,\,\,\,\,\,\,\,\,\,\,\,
\,\,\,\,\,\,\,\,\,(II)
\\SP(n,1)/(SP(n)\otimes SP(1))\,\,\,\,\,(III)\\
F_{4(-20)}/Spin(9)\,\,\,\,\,\,\,\,\,\,\,\,\,\,\,\,\,\,\,\,\,\,\,\,\,\,\,\,\,
\,\,\,\,\,\,(IV)
\end{array} \right]
\mbox{,}
\eqno{(A.3)}
$$
where $n\geq 2$. Then (see for detail Ref. \cite{will97-38-796})
$$
SO(p,q)\stackrel{def}{=}\left\{\mbox{g}\in GL(p+q,\R\left|_
{\mbox{det} \mbox{g}=1}^{\mbox{g}^tI_{pq}\mbox{g}=I_{pq}}\right\}\right.
\mbox{,}
$$
$$
SU(p,q)\stackrel{def}{=}\left\{\mbox{g}\in GL(p+q,\C)\left|_{\mbox{det}
\mbox{g}=1}
^{\mbox{g}^tI_{pq}\overline{\mbox{g}}=I_{pq}}\right\}\right.
\mbox{,}
$$
$$
SP(p,q)\stackrel{def}{=}\left\{\mbox{g}\in GL(2(p+q),\C)\left|_{\mbox{g}
^tK_{pq}\overline{\mbox{g}}=K_{pq}}^{\mbox{g}^tJ_{p+q}\mbox{g}
=J_{p+q}}\right\}\right.
\mbox{,}
\eqno{(A.4)}
$$
where $I_n$ is the identity matrix of order $n$ and 
$$
I_{pq}=\left( \begin{array}{ll}
-I_p\,\,\,\,\,\, 0\\
\,\,0\,\,\,\,\,\,\,\,\, I_q \\
\end{array} \right),\,\,\,
J_{n}=\left( \begin{array}{ll}
0\,\,\,\,\,\,\,\, \,\,\,\,I_n\\
-I_n\,\,\,\,\,\,\,0 \\
\end{array} \right),\,\,\,
K_{pq}=\left( \begin{array}{ll}
I_{pq}\,\,\,\,\,\, 0\\
\,\,0\,\,\,\,\,\,\,\,\, I_{pq} \\
\end{array} \right)
\mbox{.}
\eqno{(A.5)}
$$
The groups $SU(p,q), SP(p,q)$ are connected; the group $SO_1(p,q)$ is defined 
as the connected component of the identity in $SO(p,q)$ while $F_{4(-20)}$ is 
the unique real form of $F_4$ (with 
Dynkin diagram $\circ-\circ=\circ-\circ$) for which the character
$(\mbox{dim}X - \mbox{dim}K)$ assumes the value $(-20)$ \cite{helg62}. We 
assume that if $G=SO(m,1)$ or $SU(q,1)$ then $m$ is even 
and $q$ is odd.

Let the data $(G,K,\Ga)$ be as in Sect. 2, therefore $G$ being one of the
four groups of Eq. (A.3). The trace formula holds\cite{wall76-82-171,will90-242}
$$
\omega_{\Ga}(t;b,\chi)=V\int_{\R}dre^{-(r^2+b+\rho_0^2)t}|C(r)|^{-2}+
\theta_{\Ga}(t;b,\chi)
\mbox{,}
\eqno{(A.6)}
$$
where by definition,
$$
V\stackrel{def}{=}\frac{1}{4\pi}\chi(1)\mbox{vol}(\Ga\backslash G)
\mbox{,}
\eqno{(A.7)}
$$
where $\chi$ is a finite-dimensional unitary representation (or a character) 
of $\Ga$, and the number $\rho_0$ is associated with the positive restricted
(real) roots of $G$ (with multiplicity) with respect to a nilpotent factor 
$N$ of $G$ in an Iwasawa decomposition $G=KAN$. One has $\rho_0=(n-1)/2,\, n, 
\,2n+1,11$ in the cases $(I)-(IV)$ respectively in Eq. (A.3). Finally the 
function $\theta_{\Ga}(t;b,\chi)$ is defined as follows
$$
\theta_{\Ga}(t;b,\chi)\stackrel{def}{=}\frac{1}{\sqrt{4\pi t}}
\sum_{\ga\in C_\Ga-\{1\}}\chi(\ga)t_\ga j(\ga)^{-1}C(\ga)
e^{-(tb+t\rho_0^2+t_\ga^2/(4t))}
\mbox{,}
\eqno{(A.8)}
$$
for a function $C(\ga)$,\,\, $\ga\in\Ga$, defined on $\Ga-\{1\}$ by
$$
C(\ga)\stackrel{def}=e^{-\rho_0t_\ga}|\mbox{det}_{n_0}\left(\mbox{Ad}
(m_\ga e^{t_\ga H_0})^{-1}-1\right)|^{-1}
\mbox{.}
\eqno{(A.9)}
$$

The notation used in Eqs. (A.8) and (A.9) is the following. Let $a_0, n_0$
denote the Lie algebras of $A, N$. Since the rank of $G$ is 1, $\dim a_0=1$ 
by definition, say $a_0={\R}H_0$ for a suitable basis vector $H_0$. One can 
normalize the choice of $H_0$ by $\sigma(H_0)=1$, where $\sigma: a_0\mapsto
\R$ is the positive root which defines $n_0=\mbox{g}_{\sigma}\oplus \mbox{g}_
{2\sigma}$; for more detail see Ref. \cite{will97-38-796}. 

Since $\Ga$ is torsion free, each $\ga\in\Ga-\{1\}$ can be represented uniquely
as some power of a primitive element $\delta:\ga=\delta^{j(\ga)}$ where 
$j(\ga)\geq 1$ is an integer and $\delta$ cannot be written as $\ga_1^j$ for 
$\ga_1\in \Ga$, \,\,$j>1$ an integer. Taking $\ga\in\Ga$, $\ga\neq 1$, one 
can find $t_\ga>0$ and $m_\ga\in K$ satisfying $m_\ga a=am_\ga$ for every 
$a\in A$ such that $\ga$ is $G$ conjugate to $m_\ga\exp(t_\ga H_0)$, namely 
for some $\mbox{g}\in G, \,\mbox{g}\ga \mbox{g}^{-1}=m_\ga\exp(t_\ga H_0)$. 
For $\mbox{Ad}$ denoting the adjoint representation of $G$ on its complexified
Lie algebra, one can compute $t_\ga$ as follows \cite{wall76-82-171}
$$
e^{t_\ga}=\mbox{max}\{|c||c= \mbox{an eigenvalue of}\,\, \mbox{Ad}(\ga)\}
\mbox{,}
\eqno{(A.10)}
$$
in case $G=SO_1(m,1)$, with $|c|$ replaced by $|c|^{1/2}$ in the other cases 
of Eq. (A.3).

\s{The Spectral Zeta Function}

The zeta function $\zeta_{\Ga}(s|{\cal L})$ converges absolutely for 
$\Re s>d/2$, is holomorphic in $s$ in this domain, and for $\Re s>d/2$
$$
\zeta_{\Ga}(s|{\cal L})=\frac{\chi(1)\mbox{Vol}(\Ga\backslash G)}{4\pi}
{\cal I}(s;\alpha^2)+ \frac{1}{\Ga(s)}T_{\Ga}(s;\alpha,\chi)
\mbox{,}
\eqno{(B.1)}
$$
where $\alpha^2=b+\rho_0^2$ and \cite{will97-38-796}
$$
{\cal I}(s;\alpha^2)=\int_{\R}\frac{|C(r)|^{-2}dr}{(r^2+\alpha^2)^s}
\mbox{,}
\eqno{(B.2)}
$$

$$
T_{\Ga}(s;\alpha,\chi)=\frac{\pi^{-\frac{1}{2}}}{(2\alpha)^{s-\frac{1}{2}}}
\sum_{\ga\in C_\Ga-\{1\}}\chi(\ga)j(\ga)^{-1}C(\ga)t_{\ga}^{s+\frac{1}{2}}
K_{-s+\frac{1}{2}}(t_{\ga}\alpha)
$$

$$
=\frac{1}{\Ga(1-s)}\int_0^{\infty}\Psi_{\Ga}(t+\rho_0+\alpha;\chi)
(2\alpha t +t^2)^{-s}dt
\mbox{.}
\eqno{(B.3)}
$$

The function 
$\Psi_{\Ga}(s;\chi)$ is defined in Ref. \cite{gang77-21-1}

$$
\Psi_{\Ga}(s;\chi)=\sum_{\ga\in C_\Ga-\{1\}}\chi(\ga)t_{\ga}j(\ga)^{-1}C(\ga)
e^{-(s-\rho_0)t_{\ga}}
\mbox{,}
\eqno{(B.4)}
$$
for $\mbox{Re}s>2\rho_0$. Thus $\Psi_{\Ga}$ is a holomorphic function in the 
$\frac{1}{2}$ plane $\mbox{Re}s>2\rho_0$ and admits a meromorphic 
continuation to the full complex plane. It can be shown that 
$\Psi_{\Ga}(s;\chi)=Z_{\Ga}'(s;\chi)/Z_{\Ga}(s;\chi)$, 
where $Z_{\Ga}(s;\chi)$ is a meromorphic suitable normalized
Selberg zeta function attached to $(G,K,\Ga,\chi)$ (see Refs. 
\cite{selb56-20-47,frie77-10-133,gang77-21-1,gang80-78-1,scot80-253-177,
waka85-15-235,will92-105-163,byts96-266-1}).

The suitable Harish-Chandra-Plancherel measure is given as follows:
$$
|C(r)|^{-2}=\left[ \begin{array}{ll}C_G\pi rP(r)\tanh(\pi r), \hspace{1.0cm}
\mbox{for $G=SO_1(2n,1)$},\\
C_G\pi rP(r)\tanh(\pi r/2), \hspace{0.6cm}\mbox{for $G=SU
(q,1),\,\,\,\,\,\,\, q$ odd},\\
\hspace{4,5cm}\mbox{or $G=SP(m,1),\,\,\,\,\, F_{4(-20)}$},\\C_G\pi 
rP(r)\coth(\pi r/2), \hspace{0.7cm}\mbox{for $G=SU
(m,1),\,\,\,\,m$ even},\\
C_G\pi P(r), \hspace{2.8cm}\mbox{for $G=SO_1(2n+1,1)$},\\
\end{array} \right]
\mbox{,}
\eqno{(B.5)}
$$
while $C_{G}$ is some constant depending on $G$, and where the $P(r)$ are
even polynomials (with suitable coefficients $a_{2l}$) of degree $d-2$ for
$G\neq SO(2n+1,1)$, and of degree $d-1=2n$ for $G=SO_1(2n+1,1)$ 
\cite{byts96-266-1,will97-38-796}.

For $\mbox{Re}s>d/2$ and for $G\neq SO_1(m,1),SU(p,1)$ with $m$ odd and $p$
even we have \cite{will97-38-796} 
$$
{\cal I}(s;\alpha^2)=\frac{\pi}{2}a(G)C_G W(s;\alpha^2,a(G))
\mbox{,}
\eqno{(B.6)}
$$
where
$$
W(s;\alpha^2,a(G))= \sum_{j=0}^{\frac{d}{2}-1}a_{2j}j!
\sum_{l=0}^j
\frac{{\cal K}_{j-l}(s-l-1;\alpha^2,a(G)}
{(j-l)!(s-1)(s-2)...(s-(l+1))}
\mbox{.}
\eqno{(B.7)}
$$

For $G=SU(p,1)$ with $p$ even and $\mbox{Re}s>d/2=p$, 
$$
{\cal I}(s;\alpha^2)=C_G\pi\left[\frac{\pi}{4}W(s;\alpha^2,\frac{\pi}{2})+
\sum_{j=0}^{p-1}a_{2j}{\cal J}_j(s;\alpha^2,\frac{\pi}{2})\right]
\mbox{.}
\eqno{(B.8)}
$$

Finally for $G=SO_1(2n+1,1)$ and $\mbox{Re}s>(d/2)=(2n+1)/2$,
$$
{\cal I}(s;\alpha^2)=2C_G\pi\sum_{j=0}^n a_{2j}\int_{0}^{\infty}\frac{r^{2j}dr}
{(r^2+\alpha^2)^s}
$$

$$
=\frac{C_G\pi}{\Ga(s)}\sum_{j=0}^n
a_{2j}\alpha^{2(j+\frac{1}{2}-s)}\Ga(j+\frac{1}{2})\Ga(s-j-\frac{1}{2})
\mbox{.}
\eqno{(B.9)}
$$
In Eqs. (B.7) and (B.8) the entire functions ${\cal K}_n(s;\delta,a)$ and
${\cal J}_n(s;\delta,a)$ are defined for $\delta,a>0$ by
$$
{\cal K}_n(s;\delta,a)=\int_{{\R}}\frac{r^{2n}\mbox{sech}^{2}(ar)dr}
{(r^2+\delta^{2})^s}
\mbox{,}
\eqno{(B.10)}
$$

$$
{\cal J}_n(s;\delta,a)=\int_{{\R}}\frac{r^{2n+1}\mbox{csch}(ar)
\mbox{sech}(ar)dr}
{(r^2+\delta^2)^s}
\mbox{.}
\eqno{(B.11)}
$$

\end{document}